\documentclass[a4paper,twoside,10pt]{letter}
\usepackage{saj,multicol,subeqnarray}
\usepackage[pdftex]{graphicx}
\pdfoutput=1

\def\papertype{Invited Review}

\begin{document}
\baselineskip=3.1truemm
\columnsep=.5truecm
\newenvironment{lefteqnarray}{\arraycolsep=0pt\begin{eqnarray}}
{\end{eqnarray}\protect\aftergroup\ignorespaces}
\newenvironment{lefteqnarray*}{\arraycolsep=0pt\begin{eqnarray*}}
{\end{eqnarray*}\protect\aftergroup\ignorespaces}
\newenvironment{leftsubeqnarray}{\arraycolsep=0pt\begin{subeqnarray}}
{\end{subeqnarray}\protect\aftergroup\ignorespaces}
%


\markboth{\eightrm GREEN VALLEY} {\eightrm S. SALIM}

{\ }

\publ

\type

{\ }


\title{GREEN VALLEY GALAXIES} 


\authors{Samir Salim$^{1}$}

\vskip3mm


\address{$^1$Department of Astronomy, Indiana University,
  Bloomington, IN 47404, USA}

\Email{salims}{indiana.edu}


\dates{November 18, 2014}{November 18, 2014}


\summary{ The ``green valley'' is a wide region separating the blue and
the red peaks in the ultraviolet-optical color magnitude diagram,
first revealed using {\it GALEX} UV photometry. The term was coined by
Christopher Martin (Caltech), in 2005. Green valley highlights
the discriminating power of UV to very low relative levels of ongoing
star formation, to which the optical colors, including $u-r$, are
insensitive.  It corresponds to massive galaxies below the
star-forming, ``main'' sequence, and therefore represents a critical
tool for the study of the quenching of star formation and its possible
resurgence in otherwise quiescent galaxies. This article reviews the
results pertaining to (predominantly disk) morphology, structure,
environment, dust content and gas properties of green valley galaxies
in the local universe. Their relationship to AGN is also
discussed. Attention is given to biases emerging from defining the
``green valley'' using optical colors. We review various evolutionary
scenarios and we present evidence for a new, {\it quasi-static} view of
the green valley, in which the majority (but not all) of galaxies
currently in the green valley were only partially quenched in the
distant past and now participate in a slow cosmic decline of star
formation, which also drives down the activity on the main sequence,
presumably as a result of the dwindling accretion/cooling onto galaxy
disks. This emerging synthetic picture is based on the
findings from Fang et al.\ (2012), Salim et al.\ (2012) and Martin et
al.\ (2007), as well as other results.}



\begin{multicols}{2}
{


\section{1. INTRODUCTION}

The dichotomy between spiral and elliptical galaxies has been
established at the start of the extragalactic astronomy (Hubble
1926). The dichotomy consists of contrasts in morphology (spiral arms
vs.\ featureless), color (blue vs.\ red), kinematics (rotational vs.\
pressure supported), typical luminosity (lower vs.\ higher),
clustering (lower vs.\ higher density environments), etc. Many of
these differences are the result of, or are related to the variations in
the cold gas content, which in turn lead to very different levels of
star formation (SF). Early-type galaxies (ETGs), which include
ellipticals and disk-bearing lenticulars (S0), were traditionally
considered to represent a quiescent population, although individual
exceptions were known, particularly among S0s. The formation history
of galaxies of different morphological types remains one of the central
questions in current research efforts.

The new angle on the old question of galaxy dichotomy was provided
with the advent of the Sloan Digital Sky Survey (SDSS, York et al.\
2000). With its vast spectroscopic survey of galaxies in the local
universe ($z\sim 0.1$, Strauss et al.\ 2002) and accurate optical
photometry in five bands, SDSS enabled robust statistical analysis of
galaxy populations. A color-magnitude diagram (CMD) constructed with
SDSS photometry revealed that {\it field} galaxies formed two peaks in
their optical color distribution (Strateva et al.\ 2001, Baldry et
al.\ 2004). The narrower red peak was previously studied primarily in
galaxy clusters, and was known as the {\it red sequence}. SDSS
highlighted that the red sequence, and consequently the ETGs that are
found in it, are abundant in non-cluster environments. The wider blue
peak in optical color distribution became known as the {\it blue
cloud}. The physical differences between two populations were
thoroughly explored and quantified by Kauffmann et al.\ (2003a), who
found that the red peak galaxies have on average older stellar
populations, higher surface stellar mass densities, and dominate at
stellar masses above $10^{10.5} M_{\odot}$.

This {\it bimodality} in color distributions thus became a central
point in studies of local galaxies, but also at higher redshifts. Bell
et al.\ (2004) and Faber et al.\ (2007) reported that the luminosity
function of red-sequence galaxies has increased by a factor of two or
more since $z\sim1$. This result would suggest that the build-up of
the red sequence (and thus, presumably, of ETGs) is a process that is
ongoing at the present epoch. This scenario would be at odds with the
traditional picture in which ETGs (especially ellipticals), have
formed, and, therefore, stopped forming stars and became red, very
early in the history of the universe. This traditional picture,
supported by ample observational evidence (e.g., Trager et al.\ 2000),
has its roots in the monolithic collapse scenario of Eggen, Linden \&
Bell (1962), which has subsequently been replaced with the
hierarchical formation scenario (e.g., Kauffmann et al.\ 2006), in
which mergers of disk galaxies provide a natural formation mechanism
for elliptical galaxies (e.g., Barnes \& Hernquist 1996). If mergers
continue to be important in the latter epochs of the universe, as some
numerical simulations suggested, then they would open the door for the
late-epoch formation of ETGs and explain the reported growth of the
red sequence. And if ETGs continue to be formed today, then there
should exist galaxies that are currently in the process of
transformation. Such galaxies would have properties that are in
between those of the late and early-type galaxies. For example, they
should have some SF, but most of the former activity would have
ceased, i.e., their {\it specific star formation rate} (SSFR), or star
formation rate (SFR) per unit stellar mass (SFR/$M_*$), would be lower
than that of late-type galaxies of the same mass. The optical
bimodality recognizes only actively star-forming and quiescent
galaxies. So, how can such transitional population be identified on a
large scale?

One promising approach is to utilize {\it ultraviolet} (UV)
photometry. UV covers the peak of the blackbody emission of young
($<100$ Myr) stars, which means that even relatively small amounts of
SF will stand out in the UV, in contrast to the optical emission that
is dominated by older stars. Furthermore, the UV emission, being
produced by short-lived stars, will more closely reflect the {\it
current} SF, unlike the optical emission (even in $U$ band) which is
produced by stars that live in excess of a Gyr (Kennicutt 1998). UV
observations require observations from space. In 2003 NASA launched
{\it Galaxy Evolution Explorer}, or {\it GALEX} (Martin et al.\ 2005),
a small space telescope dedicated to mapping the sky in two
ultraviolet bandpasses: far-UV (FUV, $\lambda_{\rm eff}=1550$ \AA),
and near-UV (NUV, $\lambda_{\rm eff}=2300$ \AA). Over its mision
lifetime of ten years, {\it GALEX} observed much of the sky. Of
special significance is its medium-deep imaging survey (MIS) of
several thousand square degrees, which largely overlaps with the
SDSS. It is the synergy between {\it GALEX} with SDSS that led to the
discovery of the ``green valley''.

\centerline{\includegraphics[width=\columnwidth, keepaspectratio]{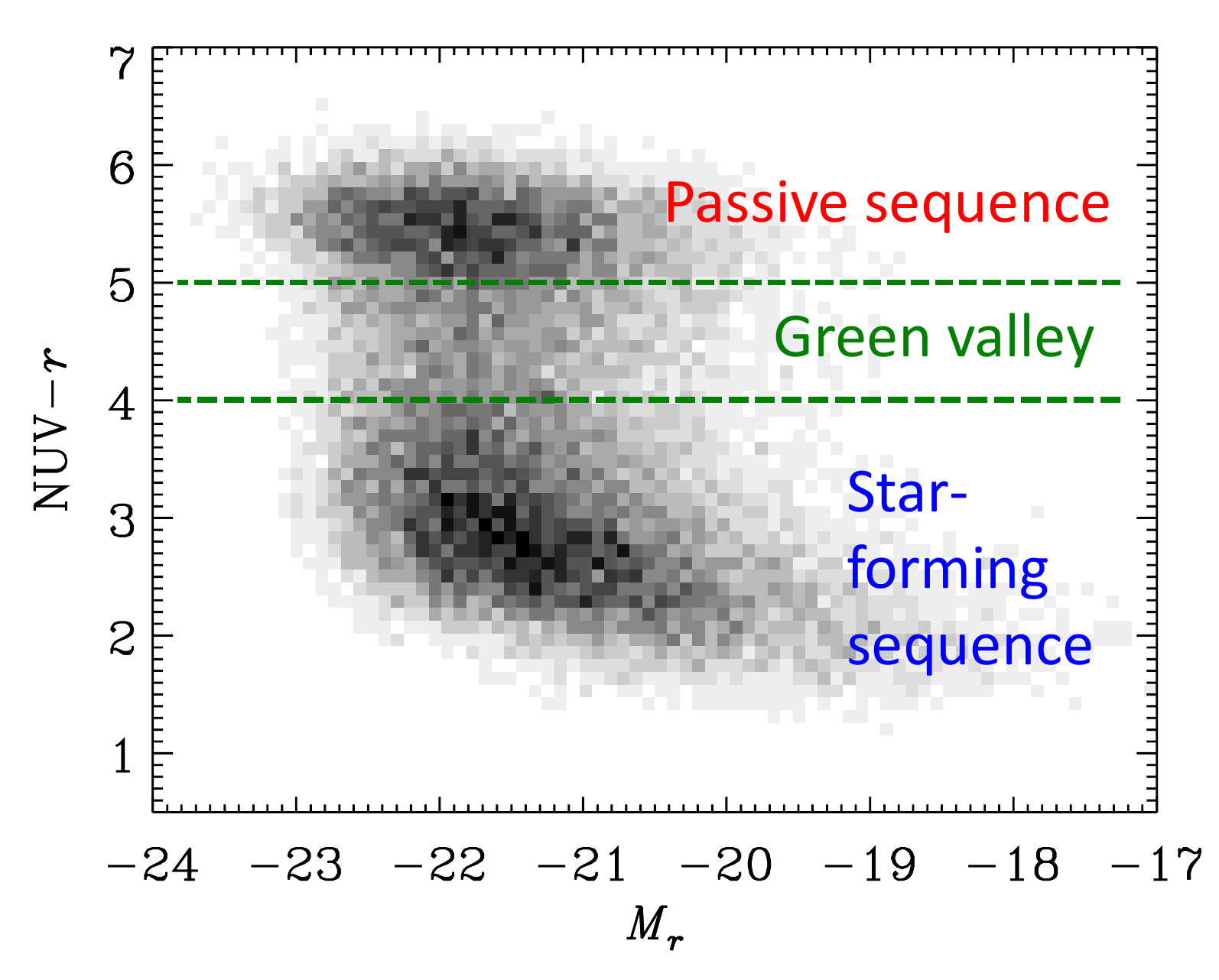}}

\figurecaption{1.}{Ultraviolet-optical color magnitude diagram showing
the ranges of NUV$-r$ colors that correspond to actively star-forming
galaxies, the green valley, and passive galaxies. This figure and
Figures 3 and 4 are based on the $z<0.22$ {\it GALEX}/SDSS sample and
data from Salim et al.\ (2007).}

\section{2. GREEN VALLEY DEFINED}

Green valley is a {\it wide} and relatively flat (hence, ``the
valley'') region in the {\it UV-optical} CMD that lies between the
peaks formed by star-forming and passive galaxies, respectively
(Figure 1).\footnote{To avoid the confusion with its usual definition
that pertains to optical colors, we will refer to red galaxies in the
UV-optical CMD as the {\it passive sequence}, rather than the red
sequence.} The term {\it green valley} was coined by D.\ Christopher
Martin, the principal investigator of {\it GALEX}, at the team meeting
held in October 2005. The green valley was initially described in a
series of {\it GALEX} papers. Wyder et al.\ (2007) defined it as a
feature of CMD; Martin et al.\ (2007) discussed its possible
evolutionary status, and with Salim et al.\ (2007) explored its
relation to active galactic nuclei (AGN).  Schiminovich et al.\ (2007)
studied its role in the morphological evolution.\footnote{Ironically,
of the four aforementioned papers that appeared in the special volume
of {\it The Astrophysical Journal} dedicated to {\it GALEX}, only
Martin et al.\ did not mention the green valley by name, presumably
because the term was considered too informal when the paper was
submitted in early 2006. The other three papers were completed later
in 2006 or 2007, by which time the term became adopted within the
team.}

\centerline{\includegraphics[width=\columnwidth, keepaspectratio]{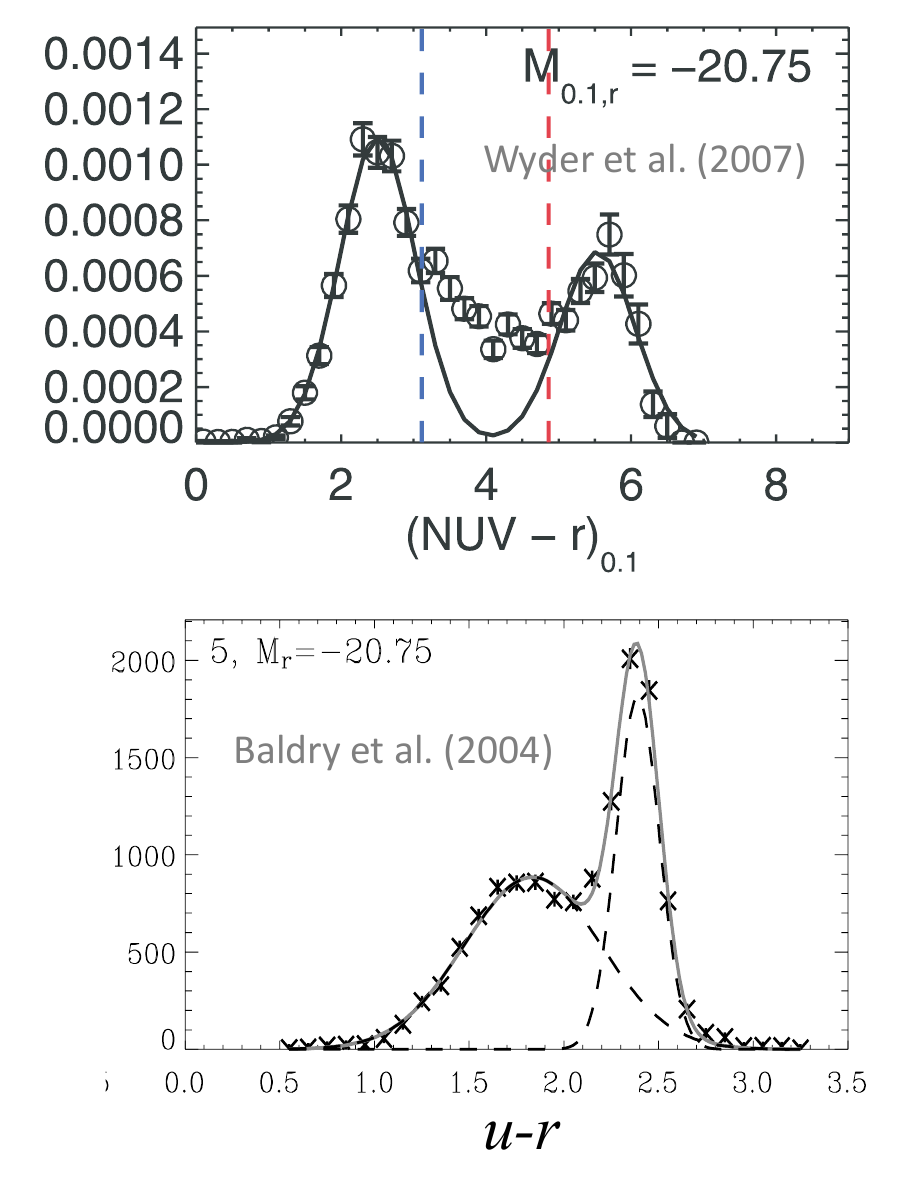}}

\figurecaption{2.}{UV-optical (NUV$-r$; upper panel) and optical
($u-r$; lower panel) color distributions of galaxies with
$M_r=-20.75$. Distributions represent slices of volume-corrected
color-magnitude diagrams. NUV$-r$ distribution has widely separated
peaks and the excess population defined as the green valley (vertical
dashed lines). In $u-r$ the blue peak is wide (the ``blue cloud'') and
overlaps with the red peak (``red sequence''), and the entire
distribution can be modeled with two Gaussians. Adapted from Wyder et
al.\ (2007) and Baldry et al.\ (2004).}

Wyder et al.\ (2007) defined the green valley as an {\it excess
population} that is left over when the rest-frame UV-optical color
distribution (in slices of optical absolute magnitude) is modeled as
the sum of two Gaussians (Figure 2a). The wide separation between the
peaks and the presence of an excess population between them stands in
stark contrast to the optical color distribution, which can be fully
decomposed, at every absolute magnitude, into a sum of two Gaussian
distributions (Baldry et al.\ 2004). This is true not only for $g-r$
color, but also for $u-r$ (Figure 2b). Wyder et al.\ (2007) pointed
out that the existence of the green valley argues for a continuum of
properties from star-forming to quiescent galaxies that is hidden
under the strict bimodality presented by optical CMDs. The green
valley is most conspicuous when rest-frame NUV magnitude (throughout
the article we refer to rest-frame magnitudes and colors) is combined
with some optical band, usually $r$ (Figure 1).

The physical significance of the green valley is that it represents a
simple tool to identify transitional galaxies.\footnote{We distinguish
between {\it transitional} and {\it transiting} galaxies, because the
latter implies an evolutionary scenario.} Transitional galaxies have
lower SSFRs than actively star-forming galaxies of the same mass
(Figure 3). The latter lie on the star-forming (``main'') sequence
(Salim et al.\ 2007, Noeske et al.\ 2007). SSFR is a proxy for the
star formation history and the evolutionary stage that the galaxy is
in. The characteristic SSFR of normal SF is mass-dependent (Figure
3). We define the onset of the transitional region to be at SSFR below
that of massive Sbc galaxies (Salim et al.\, in prep.), i.e., log SSFR
$<-10.8$ in units of Solar mass per year. Sbc's are the earliest
galaxy type in which no classical bulge is present (Fisher \& Drory
2008), so SF is expected to proceed normally (without being quenched),
and yet Sbc's extend to the massive end of the stellar mass
distribution. The lower end of the transitional range is more
difficult to determine because the accuracy of SSFRs rapidly
deteriorates below the star-forming sequence. At log SSFR $=-11.8$,
our adopted lower limit, the majority of galaxies no longer show
evidence for SF in UV images. Below this value, the SSFRs in Figures 3
and 4 should be considered upper limits because SED model libraries,
having exponentially declining SF histories with the shortest
e-folding time of 1 Gyr, do not contain models with arbitrarily low
SSFRs. This also means that the ``actual'' distribution of galaxies in
log SSFR may not have two peaks, rather, there would be the peak from
the SF sequence and the tail of galaxies that extend towards ever
lower SSFRs (Schiminovich et al.\ 2007). The bimodality seen in color
distributions is simply because red colors have a finite limit, i.e,
they saturate.

To summarize, transitional galaxies in the local universe can be
defined as:
$$
-11.8<\log ({\rm SFR}/M_*) <-10.8.
$$
\noindent Note that there is no evidence that these boundaries 
should include mass dependence for $\log M_*>9$. 

}
\end{multicols}

\centerline{\includegraphics[width=0.73\textwidth, keepaspectratio]{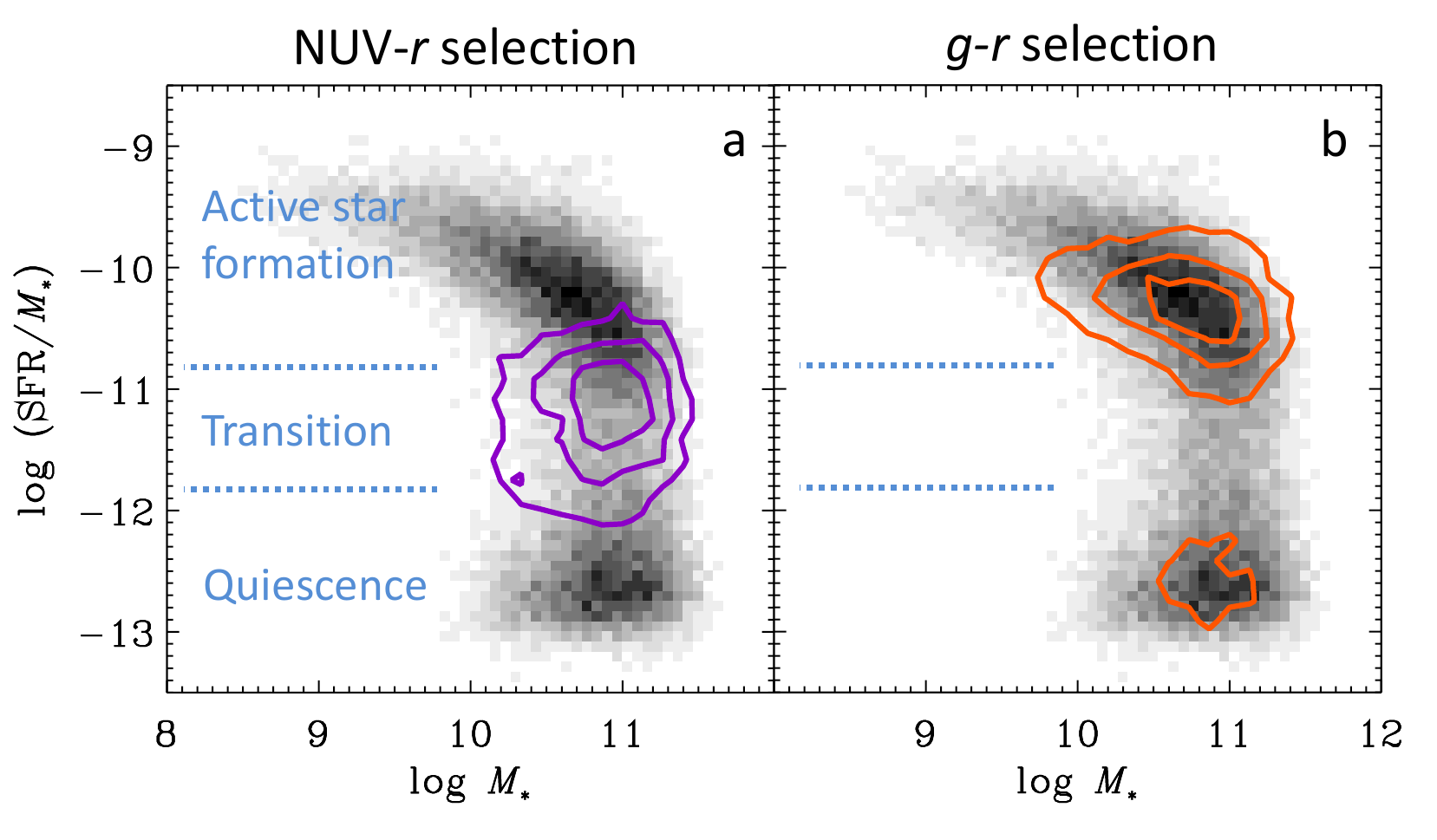}}

\figurecaption{3.}{Specific SFR vs.\ stellar mass diagram highlighting
the transition region below the star-forming sequence. Overlaid
contours show the location of galaxies selected using: (a)
intermediate NUV$-r$ colors (i.e., the green valley), and (b)
intermediate $g-r$ colors. The latter avoids the transition region and
selects massive galaxies on the ``main'' sequence. SFRs in this figure
are dust-corrected and have been determined using Bayesian SED fitting
involving up to seven bands (FUV to $z$).}

\begin{multicols}{2}
{

To illustrate the critical difference between UV-optical and optical
colors in relation to transitional galaxies, in Figure 4 we show a
variety of colors plotted against the SSFR. Figure 3b shows that
NUV$-r$, even though not corrected for dust extinction, correlates
well with SSFR over four orders of magnitude (Salim et al.\ 2005). The
green valley in NUV$-r$ can be defined as:
$$
4<{\rm NUV}-r<5.
$$

\noindent To make the above cut based on color better reflect the
transitional galaxies it is recommended to remove dusty star-forming
galaxies. This can be easily achieved by selecting face-on systems,
e.g., with a cut $b/a>0.65$ (Fang et al.\ 2012), or by introducing an
additional cut based on $r-J$ color (Bundy et al.\ 2010 modification
of Williams et al.\ 2009 method). Of course, it is best to, if
possible, completely forgo selection based on color and use
dust-corrected SSFRs instead (e.g., from SDSS MPA/JHU online catalog.)

Figure 3a shows FUV$-r$ vs.\ log SSFR. The correlation with SSFR
persists, but is somewhat less tight. More importantly, at MIS depth
{\it GALEX} FUV detection ($>3\sigma$) rate of SDSS spectroscopic
sample is only 57\% , while it is 85\% in NUV. Therefore, NUV is a
more useful choice. At a given SSFR, the spread in colors (optical or
UV-optical) of star-forming galaxies is mostly due to various amounts
of intrinsic dust reddening (and for quiescent galaxies due to
variations in stellar metallicity).\footnote{This explains why
including the absolute magnitude (with which both the dust extinction
and metallicity correlate) as the third axis in $g-r$ vs.\ NUV$-r$
diagram reduces the scatter in the color-color plane of Chilingarian
\& Zolotukhin (2011).}

The correlation between the color and the SSFR changes dramatically
when the UV band is replaced with an optical $u$ band (3550\AA, Figure
3c). Now the galaxies with transitional SSFRs have the same range of
colors as the completely quiescent galaxies. Situation with $g-r$
colors is very similar: transitional and quiescent galaxies have the
same colors, i.e, they both lie within the red sequence (Figure
3d). The insensitivity of optical colors, including $u-r$, to recent
SF has been pointed out in a number of papers (e.g., Kauffmann et al.\
2007, Fig.\ 1; Schawinski et al.\ 2007a, Fig.\ 6; Kaviraj et al.\
2007a, Fig.\ 11; Woo et al.\ 2013).

Figure 3d shows that the optical red sequence also includes galaxies
that lie at the tip of the blue cloud. That the blue cloud and the red
sequence overlap in optical colors can also be seen in Baldry et al.\
(2004) color distributions, reproduced in Figure 2b. Masters et al.\
(2010) reported a large number of spiral galaxies with red optical
colors and proposed that they were ``passive spirals'', presumably
with little or no SF. However, Cortese (2012) showed ,using UV and
mid-IR SFRs that all ``passive spirals'' are actually on the massive
end of the star-forming sequence, having SFRs no different from other
galaxies of that mass. They conclude that, especially at high masses,
optical colors cannot distinguish between actively star-forming and
passive galaxy. For the same reason it is not appropriate to quenching
fractions with optical colors, as highlighted by Woo et al.\
(2013). To quote Kauffmann et al.\ (2007), {\it ``ultraviolet
... leads to a very different accounting of which galaxies in the
local universe are truly ``red and dead""}.

\centerline{\includegraphics[width=0.82\columnwidth, keepaspectratio]{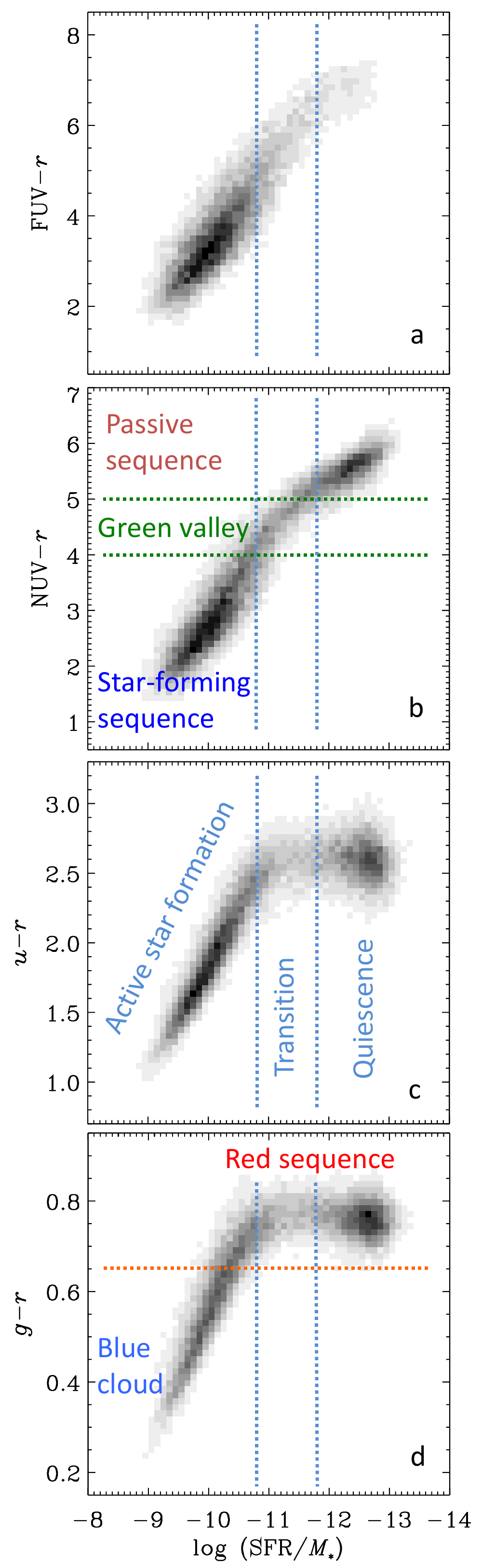}}

\figurecaption{4.}{Relationship between the rest-frame colors (no dust
correction) and dust-corrected specific SFRs. Panels a and b employ
UV-optical colors and demonstrate good correlation with the specific
SFR. In panels c and d optical colors are tightly correlated with
specific SFR for normally star-forming galaxies, but become degenerate
for low specific SFRs.}

In Figure 4a we show where the green valley galaxies selected by
UV-optical color ($4<{\rm NUV}-r<5$, no dust corrections) lie on SSFR
vs.\ stellar mass diagram (purple contours). As expected, they
correspond to galaxies that mostly lie below the star-forming
sequence. A number of studies, presumably out of convenience, define
the green valley as a region of intermediate colors in the {\it
optical CMD}. For example, Lackner \& Gunn (2012) define ``green
valley'' to be a 0.1 mag wide band in $g-r$ CMD that lies adjacent to
the red sequence. Figure 4b shows the location of these ``pseudo''
green valley galaxies on SSFR--$M_*$ diagram. It occupies the massive
end of the star-forming sequence and completely avoids the
transitional region that starts below it. It even includes some
quiescent galaxies. The reason for this ``contamination'' can be seen
in Figure 3d, where the orange dividing line corresponds to Lackner \&
Gunn split at $M_r=-22$. Some studies consider the green valley not to
be a region at all, but the {\it dividing line} (the minimum) between
the blue and the red peaks, again in the optical CMDs.

The green valley can alternatively be defined with photometric bands
other than NUV and $r$, as long as they correlate to recent SF and
stellar mass, respectively. For example, Haines et al.\ (2011) use the
ratio of 24 $\mu$m flux to $K$-band flux. Some such mid-IR equivalents
of the green valley were even given new names (Mid-infrared Canyon,
Walker et al.\ 2013; Infrared Transition Zone, Alatalo et al.\
2014). Confusion arises when mid-IR green valleys are compared, and
found to be different, from the (pseudo) ``green valley'' defined with
{\it optical} colors. A comparison with the actual green valley would
have revealed that mid-infrared colors isolate very similar
population.

In the rest of this article we will primarily discuss the results of
the studies that define the green valley using UV-optical colors, IR
colors, or utilize SSFRs directly. Furthermore, we will focus on
galaxies in the local universe ($z<0.3$) with $M>10^9 M_{\odot}$,
i.e., non-dwarf population that is more likely to consist of central
galaxies in their dark matter halos. High-redshift results are
presented separately in Section 3.6. Due to space limitations we will
omit case studies. For a more general overview of galaxy properties in
the local universe please refer to Blanton \& Moustakas (2009).

}
\end{multicols}

\centerline{\includegraphics[width=0.85\textwidth, keepaspectratio]{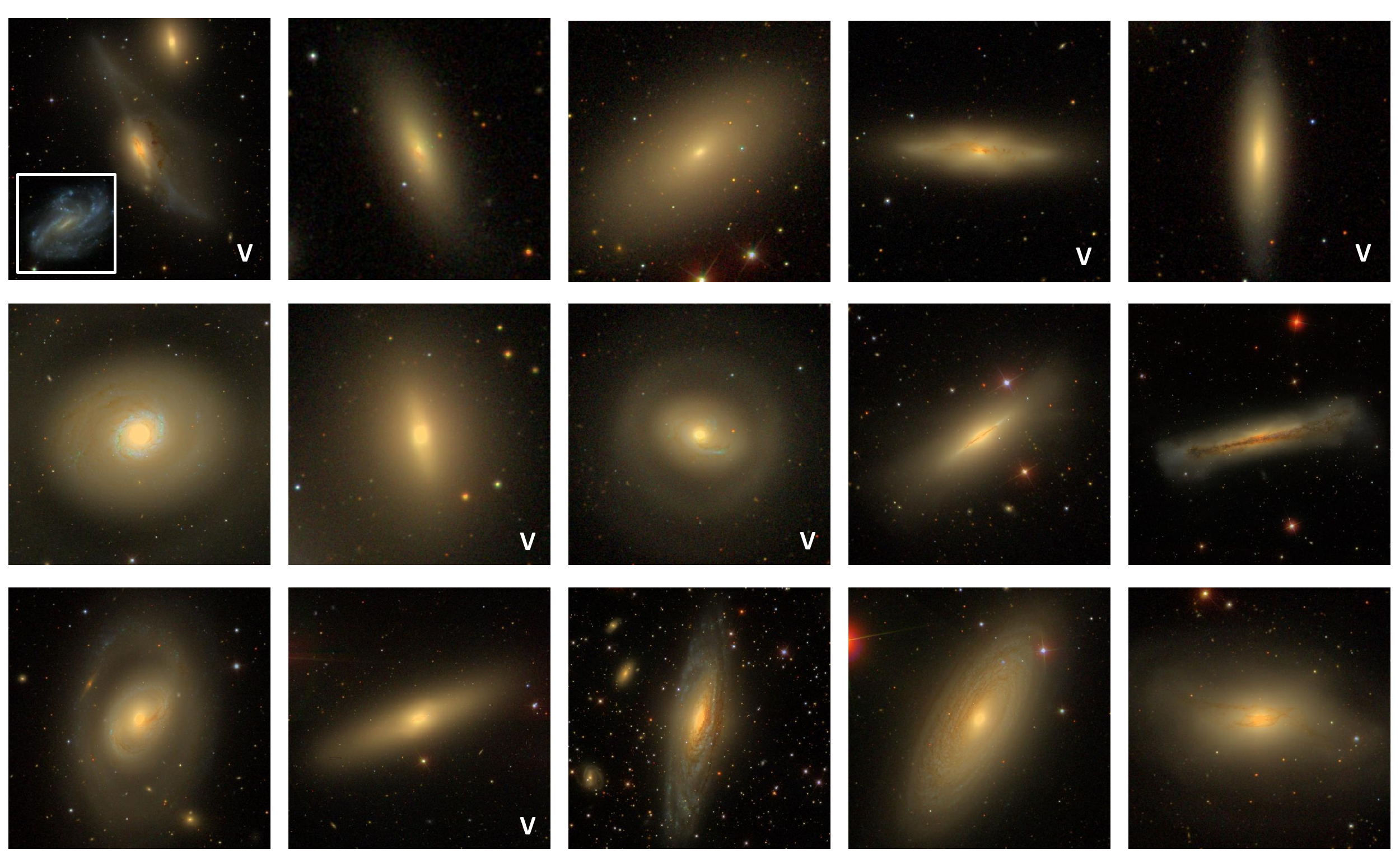}}

\figurecaption{5.}{A collage of SDSS images ($gri$ composites) of some
of the nearest ($D<25 Mpc$) green valley galaxies, spanning the mass
range from $\log M_*=9.3$ (upper left) to $\log M_*=11.4$ (lower
right). Green valley is composed of optically red, mostly disk
galaxies (S0-Sb) displaying a variety of morphological features. ``V''
denotes Virgo Cluster members. Inset shows a late-type spiral, for
color comparison. From Salim et al.\ (in prep.)}

\begin{multicols}{2}
{

\section{3. PROPERTIES OF GREEN VALLEY GALAXIES}

\subsection{3.1. Morphology and structure}

Green valley galaxies are predominantly {\it bulge-dominated, disk
galaxies}. In Figure 5 we show optical images ($gri$ composites) of a
sample of nearest ($10<D<25$ Mpc) green valley galaxies observed by
SDSS and {\it GALEX}, selected to be evenly distributed over a range
of masses ($9.3<\log M_*<11.4$). Their red appearance is consistent
with red-sequence optical color (Figure 3d). For comparison, a
lower-mass blue-cloud galaxy is shown in the inset of the first
image. Most galaxies in Figure 5 do not look like either the typical
late-type galaxies (with conspicuous and well-defined spiral arms),
nor like the typical early-type galaxies (with smooth light
profiles). Instead, they do look like a hybrid class, with often
prominent dust lanes, rings or disturbed outer profiles. Even in
galaxies that look like featureless S0s, the SF is actually present in
the disk, but requires UV imaging to be seen (Salim et al.\ 2010,
2012). Formally, the great majority of green valley galaxies are
classified as S0 through Sb in RC3 or Hyperleda catalogs. While most
S0s are are fully quiescent, some 20\% lie in the green valley. On the
other hand, Sb galaxies are primarily found on the star-forming
sequence, but some enter the green valley (Salim et al.\, in prep).
M31, an Sb galaxy with NUV$-r=4.5$ (transformed from Gil de Paz et
al.\ 2007), lies in the green valley (Mutch et al.\ 2011), but the
Milky Way (Sbc), having log SSFR$\sim=-10.5$, does not, despite
relatively red optical colors.  No single Hubble type, or even a
subtype is found only in the green valley. In this sense, the green
valley reflects the composition of the optical red sequence, in which
it is located. Red sequence includes almost all Sa's and the majority
of Sb spirals (e.g., Gil de Paz et al.\ 2007, Fig.\ 5.) {\it The red
sequence should not be equated with early-type galaxies, which in turn
must not be equated with the spheroids (elliptical galaxies).}

While pure spheroids are to be found in the green valley at higher
masses ($\log M_*>11$), there is a lack of evidence that they have
ongoing SF, i.e., they mostly scatter into the green valley from the
passive side. Salim et al.\ (2012) studied a sample of 30 green valley
galaxies with deep optical imaging from the ground and FUV follow-up
with the {\it Hubble Space Telescope (HST)}. Whenever wide-spread SF
was present in {\it HST} images (3/4 of the sample), the galaxy would
have an optical disk. They also found that the fraction of SDSS ETGs
(i.e., bulge-dominated disks and spheroids) that have extended SF (and
are therefore found in the green valley) declines above $\log
M_*>10.9$, which is where spheroids begin to dominate over disk
ETGs. The star-forming fraction also declines with optical light
concentration index, another feature of pure spheroids. Hubble
classification places pure spheroids in the {\it elliptical}
category. Consequently, ellipticals should be rare in the green
valley. This is difficult to verify with SDSS imaging which at
$z\sim0.1$ lacks the requisite resolution and depth for robust
classification, resulting in many S0s appearing like ellipticals, but
the detailed surveys of nearby ETGs (e.g., ATLAS$^{3D}$) also find
almost no evidence of SF in galaxies classified as elliptical (Young
et al.\ 2011). A notable exception is NGC 5173 (E1), the most HI-rich
known elliptical, which probably owes its patchy star-forming regions
to a particularly gas-rich merger (Vader \& Vigroux 1991). Yi et al.\
(2005) was the first to point out that as many as 15\% of ETGs lie in
the green valley and could therefore represent cases of ongoing
SF. Kaviraj et al.\ (2007a), also using {\it GALEX} photometry,
estimated that 30\% of ETGs had 1--3\% of stars younger than 1
Gyr. These result were sometimes interpreted to mean that there is a
large fraction of {\it ellipticals} with ongoing SF, but recent
results paint a more nuanced picture in which the SF in ETGs is
restricted to disk galaxies (S0s), and is largely absent from massive
ellipticals, whether they are slow or fast rotators.

It should be mentioned that the {\it UV upturn} phenomenon in ETGs
(Code \& Welch 1979), due to the old stellar populations (presumably,
horizontal branch stars) does not produce colors bluer than ${\rm
NUV}-r=5.4$ (Burstein et al.\ 1988, Schawinski et al.\ 2007a) so it
cannot, by itself, bring an ETG into a green valley. Furthermore, even
in the absence of high-resolution UV imaging, the UV upturn is
relatively easy to distinguish from SF when the UV color (i.e.,
FUV$-$NUV) is available because galaxies with the UV upturn show an
anti-correlation between UV and optical color (Donas et al.\ 2007, Gil
de Paz et al.\ 2007).

Discussion in this section does not pertain to very rare E+A galaxies,
which also have elliptical morphology and sometimes have green valley
colors (Kaviraj et al.\ 2007b). We will discuss them in Section 4.

In terms of the quantitative structural morphology, Schiminovich et
al.\ (2007) showed that the green valley galaxies have Sersic indices
that are halfway (in logarithm) between those of star-forming and
passive galaxies. Furthermore, they report that the average Sersic
indices increase with mass for every type of galaxy. Because the
Sersic index, even in $i$ band, can be affected by younger stars in
the disk (Fang et al.\ 2013), it is more physically instructive to
look at the stellar mass surface density profiles, which also place
green valley galaxies between star-forming and passive (Schiminovich
et al.\ 2007). However, when the mass surface density is considered
only in the central 1 kpc, i.e., the bulge, both the green valley and
passive galaxies have remarkably similar densities at the given mass
(Fang et al.\ 2013).

The UV morphology of green valley galaxies has been less studied than
the optical morphology, mostly because of relatively poor (5 arcsec)
resolution of {\it GALEX} images. Salim et al.\ (2012) used FUV
imaging with {\it HST}, and found that the rings of various shapes and
sizes are the most common UV feature, and are typically not
conspicuous or visible in optical images. In some cases the rings or
SF features extend well beyond the optical extent of the disk, making
these galaxies similar to giant low surface brightness galaxies such
as Malin 2 and UGC6614, or resembling extreme UV disks (Thilker et
al.\ 2007, Lemonias et l.\ 2011), that may be the signpost of intense
accretion from the intergalactic medium (IGM). In rare cases the outer
rings experience very intense SF (e.g., Hoag's object, Finkelman et
al.\ 2011), the colors of which alone would place them on the
star-forming sequence. More typically, the color profiles of
green-valley ETGs are bluer at larger radii within the contiguous
disk, with colors approaching those of actively star-forming galaxies,
but with very low surface brightnesses ($\mu_r>24$ mag arcsec$^{-1}$,
Fang et al.\ 2012), making these regions pf the disk very different
from those of actively star-forming galaxies.

\subsection{3.2. Dust and the measurement of SFRs}

Measurement of dust-corrected SFRs is quite challenging for green
valley galaxies due to their low SSFRs. Emission-line diagnostics
(H$\alpha$, [OII]) are not very useful because (a) equivalent widths
of green valley galaxies are small, (b) green valley galaxies often
contain an AGN or a LINER ($\sim 50\%$, Martin et al.\ 2007), which
contaminates line emission, (c) if spectra cover only the central
regions, as in the case of SDSS fibers, many green valley galaxies
will not have emission lines because the SF may be absent in the bulge
(Salim et al.\ 2012). Estimating SFRs and dust extinction using the
mid-infrared emission (alone, or in combination with the UV) is also
of limited use. In red sequence galaxies most of the dust heating in
the mid-IR will due to intermediate-age ($\sim$1 Gyr) or older stellar
populations (Cortese et al.\ 2008), so the estimates of SFR based on
mid-IR luminosity will be one to two orders of magnitude too high
(Salim et al.\ 2009, Hayward et al.\ 2014). Dust corrections affecting
the FUV ($A_{\rm FUV}$) can be determined from a correlation with the
UV slope ($\beta$) or UV color (Calzetti et al.\ 1994). The method
works very well for starburst galaxies (Meurer et al.\ 1999) and can
also be adapted for use with normal star-forming galaxies (e.g., Buat
et al.\ 2005, Seibert et al.\ 2005, Salim et al.\ 2007). However, the
dust-to-UV color correlation breaks down for green valley galaxies
where the UV color is primarily sensitive to the age of the stellar
population (e.g., Gil de Paz 2007). The most promising approach is to
utilize full UV-optical SED, which allows breaking of the age-dust
degeneracy (Kaviraj et al.\ 2007c). This can be achieved by fitting
the observed SED to models that include a range of dust attenuations
and SF histories. Note that the SED fitting should not extend to
rest-frame near-IR bands, since those are not well reproduced in the
current stellar population synthesis models (e.g., Taylor et al.\
2011).

It should be mentioned that the level of SF is small in green valley
galaxies only in relative terms. A $10^{11} M_{\odot}$ galaxy sitting
in the middle of the green valley (log SSFR$=-11.3$), would have a SFR
of 0.5 $M_{\odot} {\rm yr}^{-1}$, which is ten times higher than SFR
of M33 (Wilson et al.\ 1991).

Schiminovich et al.\ (2007) found that the dust attenuation of green
valley galaxies is between that of star-forming and 
truly passive galaxies. This trend is consistent with the picture in
which dust and SFR are correlated. Agius et al.\ (2013) find that
sub-mm detected ETGs tend to be found in the green valley. Also, Hinz
et al.\ (2012) report that most of the dust in an outer-ring S0 galaxy
NGC 1291 lies within the star-forming ring, which is responsible for
making it a green-valley galaxy. There have been some results that
spurred the notion that the observed green valley is composed only of
dusty SF galaxies, suggesting that the physical green valley did not
exist. That this is not be the case was pointed out already in Wyder
et al.\ (2007), who presented a dust-corrected CMD, and is obvious
from Figures 3 and 4 that do not show a gap in SSFRs.

\subsection{3.3. Environment and clustering}

Whether the reasons for a galaxy getting into the green valley are
internal or external, i.e., due to the effects of the environment, is
of critical importance. Schawinski et al.\ (2007a) studied passive and
green valley galaxies that conform to early-type morphology and have
no AGN activity. They found that ETGs tend to be somewhat bluer (in
NUV$-r$) in lower-density environments, i.e., in the field vs.\ in
groups or clusters. Also, the fraction of ETGs in the green valley was
higher for the field.. They also show that these differences are not
due to denser environments having more massive galaxies, which tend to
be more quiescent. They suggested that the change is possibly due to
ram pressure stripping in denser environments, but the results are
also consistent with other scenarios that prevent the gas from
reaching the galaxy. Crossett et al.\ (2014) found that the fraction
of red-sequence galaxies that have blue NUV$-r$ colors (i.e., scatter
into the green valley) is higher at cluster outskirts, again
consistent with environmental effects. They also indicated that
cluster green valley galaxies showed optical spiral structure, but it
was not checked if the spirals in cluster green valleys were more
common than in field green valley. They find that $\sim 10\%$ of
cluster red-sequence galaxies have some SF, which is 2--3 times lower
fraction than in the field (see Section 3.1.)

While the dependence of green-valley fractions on environment may be
suggestive of a quenching mechanism, it is not clear that the
gas-stripping by environment is sufficient to drive the galaxy towards
quiescence. Observational results are mixed. In a study of green
valley populations in Virgo cluster compared with the nearby field
sample, Hughes \& Cortese (2009) reported an intriguing result that
the Sa and (unspecified number of) {\it later} green valley galaxies
are primarily found in the cluster, but not in the field. Furthermore,
all Virgo cluster green-valley galaxies were HI deficient.  These
results suggest that the gas stripping in dense environments is
critical for galaxy quenching, but it is not clear how relevant the HI
deficiency is considering the lack of correlation with H$_2$
deficiency in clusters (Boselli et al.\ 2014).  In contrast, Haines
et al.\ (2011) report that the late-type (Sb and later) galaxies are
rare in green valleys of clusters. This forms a part of a bigger
question of the existence and the importance of ``late-type'' (small
bulge) S0s (Kormendy \& Bender 2012).

Analysis of clustering of the general population of galaxies (Heinis
et al.\ 2009, Loh et al.\ 2010) revealed that green valley galaxies
have some clustering properties similar to passive galaxies
(detachment from the Hubble flow), while most are between star-forming
and passive, suggesting that if green valley galaxies are a transiting
population, they must also change their environment.

\subsection{3.4. Active Galactic Nuclei}

Co-evolution of supermassive black holes and galaxies is one of the
central questions in studies of galaxy evolution today (Kormendy \& Ho
2013, Heckman \& Best 2014). Even relatively low-luminosity AGN may
have a role in regulating SF, or in maintaining the quiescence of a
passive galaxy (e.g., Croton et al.\ 2006).

Martin et al.\ (2007) showed that the {\it fraction} of galaxies that
are classified as AGN using the Baldwin-Peterson-Terlevich (BPT)
diagram peaks at the intermediate UV-optical colors, at $\sim50\%$,
and suggested a connection with quenching of SF, supported by an
anti-correlation, albeit weak, between AGN strength and the rate of
quenching. Salim et al.\ (2007) extended this analysis by considering
BPT AGN and non-AGN at the same stellar mass, and found that AGN, both
the raw numbers and {\it number densities} actually peak in the
star-forming sequence, with the tail towards the green
valley. Furthermore, they emphasized the difference between strong AGN
($L[{\rm OIII}]>10^7L_{\odot}$), which lie entirely on the massive end
of the star-forming sequence, and the weak AGN, which are mostly found
on the star-forming sequence but also extend to the passive sequence
(Figure 6). They present a picture in which AGN may be responsible for
a {\it gradual} suppression of SF of galaxies on the star-forming
sequence, and argue against sporadically fed AGN through minor
gas-rich mergers. The result of Martin et al., that the AGN fractions
peak in the green valley, can be reconciled with their numbers peaking
on the SF sequence by the fact that in the green valley the BPT method
is sensitive to LINERs, which may not represent active nuclei, but may
arise from excitations from hot evolved populations (Stasinska et al.\
2008).

Martin et al.\ (2007) results are sometimes misinterpreted to mean
that AGN {\it numbers} peak in the green valley, spurring criticism
that for AGN quenching to be viable they should not peak in the green
valley (e.g., Schawinski et al.\ 2009). The confusion regarding
whether AGN are found among the actively star-forming galaxies was
further compounded because there are numerous studies that define the
``green valley'' using {\it optical} colors (e.g., Nandra et al.\
2007, Silverman et al.\ 2008, Schawinski et al.\ 2007b, 2009, 2010,
Treister et al.\ 2009, Vasudevan et al.\ 2009, Cardamone et al.\ 2010,
Mendez et al.\ 2013, Smolcic 2009), and which consequently find that
AGN peak at intermediate colors (especially the more luminous ones
selected by X rays or the mid-IR slope). However, massive, normal
star-forming galaxies intrinsically can only have intermediate optical
colors (Figure 4b), so these studies actually confirm that AGN are
mostly on the star-forming sequence. The conclusion is that, when
interpreted correctly, both the UV and the optical studies present
consistent evidence (see also Rosario et al.\ 2013).

\centerline{\includegraphics[width=\columnwidth,keepaspectratio]{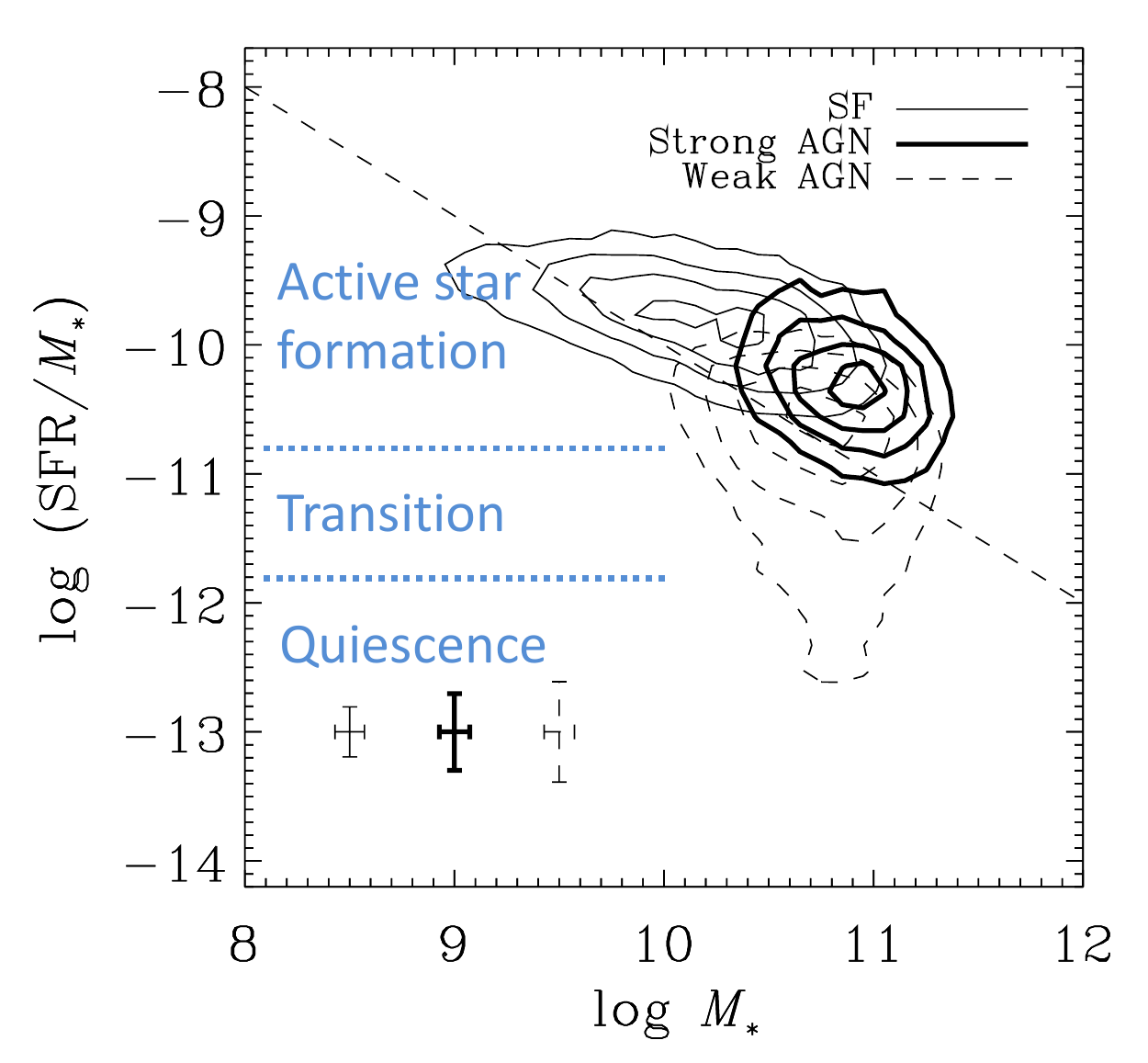}} 

\figurecaption{6.}{Specific SFR vs.\ stellar mass diagram showing the
  locations of galaxies selected from the BPT diagram to be
  star-forming (thick solid contours), strong AGN ($L{\rm [OIII]}>10^7
  L_{\odot}$) and weak AGN ($L{\rm [OIII]}<10^7 L_{\odot}$). AGN are
selected to lie above the Kauffmann et al.\ (2003b) demarkation line
on the BPT diagram. Strong BPT AGN lie on the star-forming sequence,
as well as many weak ones. The latter extend through the transition
region (the green valley). Adapted from Salim et al.\ (2007)}

It is important to point out that the BPT selection of AGN in the
local universe needs to include all galaxies lying above the empirical
demarcation line of Kauffmann et al.\ (2003b), and not only galaxies
that lie above the Kewley et al.\ (2001) line of maximum theoretical
SF (Juneau et al.\ 2014). Selecting only the latter biases the sample
towards low-luminosity AGN (Fig.\ 2 in Kauffmann et al.\ 2003b), which
on average have lower SSFRs (Kewley et al.\ 2006). AGN that lie
between the two lines are often called SF/AGN composites, but it
should be kept in mind that all galaxies on the AGN branch (AGN mixing
sequence) can have SF to some extent, and that being above the Kewley
et al.\ line is irrelevant.

That BPT identifies AGN with specific SFRs as high as those of non-AGN
(at a given mass) argues against significant incompleteness or bias of
the BPT technique (above some threshold of AGN luminosity) due to SF
contamination (see also Fig.\ 6 in Kauffmann et al.\ 2003b). This
claim is further supported by the fact that AGN found by other methods
(e.g., X-rays, radio), also fall within the AGN branch of the BPT
diagram (Yan et al.\ 2011, Juneau et al.\ 2011).

Most AGN studies in the context of the green valley population pertain
to the more common Type 2 AGN, whose central engine is not
sufficiently strong to affect the global color of the host, or is so
oriented that it is sufficiently obscured by the dust. The vast
majority of AGN belong to Type 2, even if LINERs are not considered to
be AGN. Type 1 AGN (QSOs, Seyfert 1s) on the other hand have broad
emission lines and contribute to continuum emission. Trump et al.\
(2013) remove the central sources of broad-line AGN to study host
colors, and find that such AGN prefer optically bluer hosts, in line
with the results for luminous Type 2 AGN.

\subsection{3.5. Gas properties}


Gas content of green valley galaxies, both atomic and molecular, and
its relation to galaxies on either side of the green valley provides
important clues to understanding the origin of this population. {\it
GALEX} Arecibo SDSS Survey (GASS; Catinella et al.\ 2010) obtained HI
measurements for $\sim 1000$ galaxies with $0.025<z<0.05$ and
$M_*>10^{10} M_{\odot}$, with special emphasis on transitional
galaxies. Producing a survey of HI that would not be biased towards
gas-rich systems is challenging, and GASS attempts to reduce the
effect of these biases by careful selection and analysis. Unlike
galaxies on the passive sequence, almost all green valley galaxies are
detected in GASS. Gas fraction ($M_{\rm HI}/M_*$) correlates with
NUV$-r$ color (Catinella et al.\ 2012), suggesting that the
green-valley phase is not due to the abrupt change in the gas
reservoir. Remarkably, GASS showed that the SF efficiency is nearly
the same in the green valley and outside of it (Schiminovich et al.\
2010), suggesting that whatever mechanism regulates the gas supply
also regulates the SF.

Fabello et al.\ (2011) find that the SF efficiency based on stacked HI
data from ALFALFA survey does not seem to depend on the bulge
fraction, thus disfavoring the morphological quenching scenario of
Martig et al.\ (2009). One should keep in mind that the model of
Martig et al.\ (2009) in which the bulge suppresses SF without
removing the gas is strictly applicable only to inner-disk {\it
molecular} gas, and Crocker et al.\ (2011) do find that some ETGs with
molecular gas do not have SF (based on their UV-optical color). More
generally, however, the H$_2$ fraction, like the HI fraction, does
correlate with NUV$-r$ color (Saintonge et al.\ 2011), so it is
difficult to claim that morphology regulates SF without also affecting
the gas content.

\subsection{3.6. Green valley at higher redshifts}

Because the green valley is defined relative to the star-forming
sequence (Section 2), which is shifting towards higher SSFRs with
redshift (e.g., Speagle et al.\ 2014), it is appropriate to also have
redshift-dependent cuts on SSFR and/or UV-optical color. For $\langle
z\rangle\sim0.8$ sample, Salim et al.\ (2009) use gren valley limits
of $-11<\log (SFR/M_*)<-10$ and $3.5<{\rm NUV}-r<4.5$, which match
very well with the Moustakas et al.\ (2013) division between SF and
passive galaxies that shifts in log SSFR as $1.33z$.

Bundy et al.\ (2010) followed the morphology of the green valley and
passive sequence to $z\sim 1.2$ and reported that the fraction of disk
galaxies, especially at higher masses, was greater in the past,
suggesting that some of them merge into spheroids while being
green/passive. Mendez et al.\ (2011) found, using quantitative
morphological parameters (CAS, bulge-to-total ratio, Gini/M$_{20}$)
that the green valley at $0.4<z<1.2$, is mostly composed of disk
galaxies intermediate between star-forming and passive galaxies, with
even lower merger fractions than those measured for blue galaxies,
suggesting that the mergers are not important for quenching at those
redshifts. These results do not differ essentially from those in the
local universe.

Goncalves et al.\ (2012) obtained spectra of green valley galaxies at
$z\sim0.8$ which suggest that the transiting rate used to be higher in
the past, and involved more massive galaxies, i.e., that the massive
end of the passive sequence was put in place earlier, representing
another aspect of ``downsizing''.

E+A galaxies, which represent an interesting green-valley population
locally (Section 4), have also been identified at $0.5<z<1.2$, and
were found to reside in dense environments (like quiescent
galaxies). Their number densities, however, are comparable to those of
E+As at low redshift (Vergani et al.\ 2010).

\section{4. EVOLUTIONARY PICTURE}

In this section we overview various evolutionary scenarios for the
green valley, and propose the ``quasi-static'' model as a synthesis
view.

Green valley (modulo dust) is most often portrayed as the population
of galaxies transiting, at the present epoch, from active SF to
quiescence on a relatively fast timescale ($<1$ Gyr), i.e, as galaxies
that have {\it recently} been quenched or are undergoing
quenching. The picture that has emerged over the last seven years is
both different and more complex.

From the methodological point of view it should be said that the
studies that discuss the migration of galaxies through the {\it
optical} CMD will be of limited use for understanding the quenching
mechanisms in massive galaxies. This is because the different
locations within the (dust-corrected) blue cloud simply represent
various levels of SSFR of galaxies that are actively star-forming (see
Figure 3d), with many of the massive star-forming galaxies already in
the red sequence. In other words, the transition to quiescence for
massive (non-satellite) galaxies happens {\it within} the optical red
sequence, ``under the radar'' of optical CMDs. Optical colors are
instead useful for the study of normal, pre-quenching SF, as evidenced
by the tight correlation between optical colors and SSFR (Figure 3c, d).


Before evaluating various scenarios for the origin of green valley
population, we assess the evidence that (a) there should exist a
substantial transit from active SF to quiescence, and (b) that this
transit should produce spheroids (ellipticals). Recent results show
that when the evolution of the mass function of star-forming and
quiescent galaxies is determined using the actual SSFRs and from areas
that minimize cosmic variance, there is essentially no evolution in
the massive end ($M_*>10^{10.5} M_{\odot}$) of quiescent galaxies
since $z\sim0.8$ (Moustakas et al.\ 2013). This implies that
ellipticals, which dominate at these masses, have already been in
place at that epoch, which is consistent with numerous studies that
fail to find sufficient number of major mergers at the present epoch
to sustain the formation of new ellipticals (e.g., Lotz,\ 2008), and
with the classical picture of ellipticals as an ancient
population. Thus, the flux of galaxies through the green valley may be
much smaller than previously thought based on the optical red sequence
luminosity functions, and may be limited to lower mass galaxies.

An often laid out argument for the rapid ($<1$ Gyr) transition through
the green valley comes from the relative paucity of galaxies that are found
there. However, the green-valley crossing time is impossible to
determine based on the number density of galaxies alone. 
Martin et al.\ (2007) used SF history-sensitive
spectral indices H$\delta_A$ and $D_n$(4000) to roughly determine the
quenching rates in the green valley and have found (their Figure 13)
that the quenching timescales span the entire range probed, from some with
just 50 Myr to the majority (50\%) with 2 Gyr. This latter timescale
corresponds to green-valley crossing times of at least 2.5 Gyr. They
estimate that the total transiting time from activity to
quiescence is as long as $\sim 6$ Gyr (their Section 6.1.2). Note that
this does not mean that most transiting galaxies move slowly, only
that the slow ones will be overrepresented in the green valley:
{\it ``Galaxies with faster quench rates will
spend less time in the transition region, and the fact that the
majority of galaxies are found to have low quench rates in the
transition zone does not mean that the majority of galaxies undergo
slow quenches.''} (Martin et al.\ 2007.)

That the green valley may be mostly {\it static} was proposed in Salim
et al.\ (2012) and Fang et al.\ (2012) on the basis of the detailed
investigation of the UV morphologies and star-formation histories of a
sample of green valley ETGs. Morphologies were found to be consistent
with the smooth, gradual star formation. Most SF was found to take
place in the outer regions of optically red, old disks. SF histories
of these outer disks, do not differ from those of actively
star-forming galaxies, i.e., there is no evidence of a fast decline or
fast rise. Note that in this quasi-static scenario there is no reason
for SFRs of green valley galaxies to stay constant over cosmic
epochs. They may still decline slowly for the same reasons that drive
the decline of SF on the main sequence, e.g., the drop in the overall
accretion rate (e.g., van de Voort et al.\ 2011). {\it Therefore,
most (but certainly not all) galaxies currently in the green valley
are not moving rapidly through it, rather, it is the green valley (as
the region below the star-forming sequence) that moves slowly towards
lower SSFRs.}


Suggestions have been made that some green valley galaxies may come
{\it from} the passive sequence, following the resumption of (steady)
gas accretion from the IGM (Thilker et al.\ 2010). Salim et al.\
(2012) append this scenario by suggesting that the galaxy may move
between low-level SF and quiescence depending on the duty cycle of a
(Type 2) AGN. The connection between the fueling of SF and of AGN has
been shown in Kauffmann et al.\ (2007), who exploited the fact that
SDSS fibers capture the bulge region of the galaxy, while the
photometry is global, to explore age gradients in bulge-dominated
green valley galaxies. They found that what leads to intermediate
UV-optical colors is the SF in the extended disk, whereas the bulges
can be young or old. On the other hand, old disks rarely have young
bulges. This suggests that the gas source is an outer reservoir,
rather than the gas from a merging galaxy, which usually ends up in
the centers of the galaxies (e.g., Peirani et al.\ 2010), though
interactions may help bring the gas there. {\it The outer reservoir
fuels disk SF and occasionally also reaches the central black
hole. The resulting activity may lead to the prevention of further
accretion by some feedback process and result in (temporary) shutting
down of SF}. This connection between the green valley and the passive
sequence is supported by Fang et al.\ (2013) result that the two
groups of galaxies are structurally very similar. Depending on the
duration of AGN/SF cycles and the prevalence of feedback, this
scenario may for all practical purposes be observationally
indistinguishable from the more general static picture.

In contrast to this relatively steady-state picture, a scenario has
been proposed by Kaviraj et al.\ (2009) that the green valley is made
up entirely of otherwise passive galaxies that recently underwent a
minor, gas-rich merger. They support this scenario with numerical
simulations of merging populations. While there is little doubt that
minor mergers involving quiescent galaxies and gas-rich galaxies
occur, and may not be uncommon (e.g., upper left image in Figure 5, or
Crockett et al.\ 2011), it is unlikely, from the morphologies and SF
histories, that such cases are responsible for the majority of the
green valley.

The idea of an external source of gas for green valley galaxies has
been challenged on the account of the measurements of the metal
abundance, which for some star-forming rings around S0 galaxies is
close to solar (Bresolin 2013, Ilyina, et al.\ 2014). At face value
these results argue against accretion from IGM, or at least the
accretion of pristine gas. However, in the quasi-static scenario the
accretion is a slow process, during which substantial enrichment may
be possible. The main alternative to the fueling of green valley SF
through the accretion of external gas is that it is sustained by
internal gas from mass loss. It dates back to Faber \& Gallagher
(1976), when the ubiquity of gas in and around ETGs was not well
established. Kaviraj et al.\ (2007a) claimed that the recycled gas
from mass loss is insufficient to produce SF observed in green valley
galaxies, however, it may provide some fuel (Leitner \& Kravtsov
2011). Since the mass loss should follow stellar light, and the SF in
green valley mostly happens in the outer disk, this avenue is
plausible if the gas is transported, e.g., by secular action of a bar
(Schwarz 1984). Since mass loss is universal, the question arises as
to why are not all barred S0 galaxies star-forming?

Detailed discussion of the processes that initiated the quenching and
brought the galaxy into the green valley in the first place is outside
of the scope of this review. Green valley galaxies cannot simply be
more quiescent versions of star-forming galaxies, because their
structure is different (more centrally concentrated) than that of
galaxies on the main sequence of the same mass (Schiminovich et al.\
2007, Fang et al.\ 2013). Therefore, the simple fading mechanisms, due
to, for example, gas exhaustion or gas starvation cannot explain why
galaxies leave the star-forming sequence. Recent work places
importance to the build up of the bulge (e.g., Bell 2008, Cheung et al.\ 2012),
rather than the so called halo quenching (Dekel \& Birnboim 2006, Woo
et al.\ 2013). This makes both the ejective AGN feedback (e.g.,
Cattaneo et al.\ 2006) and the morphological quenching (Martig et al.\
2009) viable options. {\it The quasi-static model of the green valley
informs us that after it was initiated, the quenching does not always
proceed to full quiescence, and therefore many of the present-day
green valley galaxies may have been quenched, but only partially,
early in the history of the universe.}

Finally, we touch upon E+A and related galaxies, which may represent
the only true fast-transiting population. E+As (Dressler \& Gunn 1983,
Yang et al.\ 2008) are characterized by the lack of [OII] and Balmer
emission lines, i.e., no current SF, but with deep H$\delta$ features,
signaling the presence of relatively young A stars. They must have
experienced a strong burst of SF in which a significant fraction of
mass has formed over a short period of time ($<0.1$ Gyr; Kaviraj et
al.\ 2007b), possibly induced by a major gas-rich merger. As a
result, E+As retain their very blue optical colors ($0.2<g-r<0.5$),
but due to the lack of {\it current} SF lie offset in NUV$-r$ colors
towards the red, with many (but not all) of E+As lying in the green
valley. They occupy a region to the right of the star-forming sequence
in Figure 3d (see also Kaviraj et al.\ 2007b). Chilingarian \&
Zolotukhin (2012) show tracks of {\it truncated} SF histories that
pass through similar regions as those occupied by E+As. Optically blue
ETGs of Kannappan et al.\ (2009) and McIntosh et al.\ (2013) may also
represent a population with truncated SF histories, but without having
experienced the burst prior to the cessation of SF. E+As and related
galaxies therefore represent the true, fast-transiting green-valley
population, but they are very rare compared to ``normal'' green valley
galaxies and not representative of it as a whole. However, due to
their fast transit times (Chilingarian \& Zolotukhin 2011), they may
be an important source of quiescent galaxies over cosmic times (Yesuf
et al.\ 2014), possibly sufficient to explain the moderate low-mass
passive sequence buildup, which the masses of E+As match well.

Green valley represents an exciting development made possible by
wide-area UV surveys. The ability to identify and study galaxies with
low, but non-zero specific SFRs has already led to important advances in
our understanding of galaxy evolution.


\acknowledgements{I acknowledge my {\it GALEX} and AEGIS collaborators
(R.\ Michael\ Rich, Stephane Charlot, Sandra\ M.\ Faber, Mark\
Seibert) and the {\it GALEX} team for their contributions over the
years. I also gratefully acknowledge the efforts of the SDSS
collaboration. I thank Cameron Pace for help with preparing the
bibliography, and the editor Dejan Uro\v{s}evi\'{c} for giving me this
wonderful opportunity.}


\references
Agius N. K., et al.: 2013, \journal{Mon. Not. R. Astron. Soc.}, {\bf 431}, 1929.  

Alatalo K., Cales S. L., Appleton P. N., Kewley L. J., Lacy M., Lisenfeld U., Nyland K., Rich J. A.: 2014, \journal{Astrophys. J.}, {\bf 794}, LL13.  

Baldry I. K., Glazebrook K., Brinkmann J., Ivezi\'{c}, \^{Z}., Lupton R. H., Nichol R. C., Szalay A. S.: 2004, \journal{Astrophys. J.}, {\bf 600}, 681.  

Barnes J. E., Hernquist L.: 1996, \journal{Astrophys. J.}, {\bf 471}, 115.  

Bell E. F., et al.: 2004, \journal{Astrophys. J.}, {\bf 608}, 752.  

Bell E. F., et al.: 2004, \journal{Astrophys. J.}, {\bf 608}, 752.  

Bell E. F.: 2008, \journal{Astrophys. J.}, {\bf 682}, 355.  

Blanton M. R., Moustakas J.: 2009, \journal{Annu. Rev. Astron. Astrophys.}, {\bf 47}, 159.  

Boselli A., Cortese L., Boquien M., Boissier S., Catinella B., Gavazzi G., Lagos C., Saintonge A.: 2014, \journal{Astron. Astrophys}, {\bf 564}, AA67. 

Bresolin F.: 2013, \journal{Astrophys. J.}, {\bf 772}, LL23.  

Buat V., et al.: 2005, \journal{Astrophys. J.}, {\bf 619}, L51.  

Bundy K., et al.: 2010, \journal{Astrophys. J.}, {\bf 719}, 1969.  

Bundy K., et al.: 2010, \journal{Astrophys. J.}, {\bf 719}, 1969.  

Burstein D., Bertola F., Buson L. M., Faber S. M., Lauer T. R.: 1988, \journal{Astrophys. J.}, {\bf 328}, 440.  

Calzetti D., Kinney A. L., Storchi-Bergmann T.: 1994, \journal{Astrophys. J.}, {\bf 429}, 582. 

Cardamone C. N., Urry C. M., Schawinski K., Treister E., Brammer G., Gawiser E.: 2010, \journal{Astrophys. J.}, {\bf 721}, L38.  

Catinella B., et al.: 2010, \journal{Mon. Not. R. Astron. Soc.}, {\bf 403}, 683. 

Catinella B., et al.: 2012, \journal{Astron. Astrophys}, {\bf 544}, AA65. 

Cattaneo A., Dekel A., Devriendt J., Guiderdoni B., Blaizot J.: 2006,  

Cheung E., et al.: 2012, \journal{Astrophys. J.}, {\bf 760}, 131.  

Chilingarian I. V., Zolotukhin I. Y.: 2012, \journal{Mon. Not. R. Astron. Soc.}, {\bf 419}, 1727.  

Code A. D., Welch G. A.: 1979, \journal{Astrophys. J.}, {\bf 228}, 95.  

Cortese L., Boselli A., Franzetti P., Decarli R., Gavazzi G., Boissier S., Buat V.: 2008, \journal{Mon. Not. R. Astron. Soc.}, {\bf 386}, 1157.  

Cortese L.: 2012, \journal{Astron. Astrophys}, {\bf 543}, AA132.  

Crocker A. F., Bureau M., Young L. M., Combes F.: 2011, \journal{Mon. Not. R. Astron. Soc.}, {\bf 410}, 1197.  

Crockett R. M., et al.: 2011, \journal{Astrophys. J.}, {\bf 727}, 115.  

Crossett J. P., Pimbblet K. A., Stott J. P., Jones D. H.: 2014, \journal{Mon. Not. R. Astron. Soc.}, {\bf 437}, 2521.  

Croton D. J., et al.: 2006, \journal{Mon. Not. R. Astron. Soc.}, {\bf 365}, 11.  

Dekel A., Birnboim Y.: 2006, \journal{Mon. Not. R. Astron. Soc.}, {\bf 368}, 2.  

Donas J., et al.: 2007, \journal{Astrophys. J. Suppl. Series}, {\bf 173}, 597. 

Dressler A., Gunn J. E.: 1983, \journal{Astrophys. J.}, {\bf 270}, 7.  

Eggen O. J., Lynden-Bell D., Sandage A. R.: 1962, \journal{Astrophys. J.}, {\bf 136}, 748.  

Fabello S., Catinella B., Giovanelli R., Kauffmann G., Haynes M. P., Heckman T. M., Schiminovich D.: 2011, \journal{Mon. Not. R. Astron. Soc.}, {\bf 411}, 993.  

Faber S. M., et al.: 2007, \journal{Astrophys. J.}, {\bf 665}, 265.  

Faber S. M., Gallagher J. S.: 1976, \journal{Astrophys. J.}, {\bf 204}, 365.  

Fang J. J., Faber S. M., Koo D. C., Dekel A.: 2013, \journal{Astrophys. J.}, {\bf 776}, 63.  

Fang J. J., Faber S. M., Salim S., Graves G. J., Rich R. M.: 2012, \journal{Astrophys. J.}, {\bf 761}, 23.  

Finkelman I., Moiseev A., Brosch N., Katkov I.: 2011, \journal{Mon. Not. R. Astron. Soc.}, {\bf 418}, 1834.  

Fisher D. B., Drory N.: 2008, \journal{Astron. J.}, {\bf 136}, 773.  

Gil de Paz A.,et al.: 2007, \journal{Astrophys. J. Suppl. Series}, {\bf 173}, 185.  

Gon\c{c}alves T. S., Martin D. C., Men\'{e}ndez-Delmestre K., Wyder T. K., Koekemoer A.: 2012, \journal{Astrophys. J.}, {\bf 759}, 67.  

Haines C. P., Busarello G., Merluzzi P., Smith R. J., Raychaudhury S., Mercurio A., Smith G. P.: 2011, \journal{Mon. Not. R. Astron. Soc.}, {\bf 412}, 145. 

Haines C. P., et al.: 2011, \journal{Mon. Not. R. Astron. Soc.}, {\bf 417}, 2831.  

Heckman T. M., Best P. N.: 2014, \journal{Annu. Rev. Astron. Astrophys.}, {\bf 52}, 589.  

Heinis S., et al.: 2009, \journal{Astrophys. J.}, {\bf 698}, 1838. 

Hinz J. L., et al.: 2012, \journal{Astrophys. J.}, {\bf 756}, 75.  

Hubble E. P.: 1926, \journal{Astrophys. J.}, {\bf 64}, 321.  

Hughes T. M., Cortese L.: 2009, \journal{Mon. Not. R. Astron. Soc.}, {\bf 396}, L41. 

Ilyina M. A., Sil'chenko O. K., Afanasiev V. L.: 2014, \journal{Mon. Not. R. Astron. Soc.}, {\bf 439}, 334.  

Juneau S., Dickinson M., Alexander D. M., Salim S.: 2011, \journal{Astrophys. J.}, {\bf 736}, 104.  

Juneau S., et al.: 2014, \journal{Astrophys. J.}, {\bf 788}, 88.   

Kannappan S. J., Guie J. M., Baker A. J.: 2009, \journal{Astron. J.}, {\bf 138}, 579.  

Kauffmann G., et al.: 2003, \journal{Mon. Not. R. Astron. Soc.}, {\bf 341}, 33.   

Kauffmann G., et al.: 2003, \journal{Mon. Not. R. Astron. Soc.}, {\bf 341}, 54.  

Kauffmann G., et al.: 2007, \journal{Astrophys. J. Suppl. Series}, {\bf 173}, 357.  

Kauffmann G., et al.: 2007, \journal{Astrophys. J. Suppl. Series}, {\bf 173}, 357.  

Kauffmann G., Heckman T. M., De Lucia G., Brinchmann J., Charlot S., Tremonti C., White S. D. M., Brinkmann J.: 2006, \journal{Mon. Not. R. Astron. Soc.}, {\bf 367}, 1394.  

Kaviraj S., et al.: 2007a, \journal{Astrophys. J. Suppl. Series}, {\bf 173}, 619.  

Kaviraj S., Kirkby L. A., Silk J., Sarzi M.: 2007b, \journal{Mon. Not. R. Astron. Soc.}, {\bf 382}, 960.  

Kaviraj S., Peirani S., Khochfar S., Silk J., Kay S.: 2009, \journal{Mon. Not. R. Astron. Soc.}, {\bf 394}, 1713.  

Kaviraj S., Rey S.-C., Rich R. M., Yoon S.-J., Yi S. K.: 2007c, \journal{Mon. Not. R. Astron. Soc.}, {\bf 381}, L74.  

Kennicutt R. C., Jr.: 1998, \journal{Annu. Rev. Astron. Astrophys.}, {\bf 36}, 189.  

Kewley L. J., Dopita M. A., Sutherland R. S., Heisler C. A., Trevena J.: 2001, \journal{Astrophys. J.}, {\bf 556}, 121.  

Kewley L. J., Groves B., Kauffmann G., Heckman T.: 2006, \journal{Mon. Not. R. Astron. Soc.}, {\bf 372}, 961.  

Kormendy J., Bender R.: 2012, \journal{Astrophys. J. Suppl. Series}, {\bf 198}, 2. 

Kormendy J., Ho L. C.: 2013, \journal{Annu. Rev. Astron. Astrophys.}, {\bf 51}, 511.  

Lackner C. N., Gunn J. E.: 2012, \journal{Mon. Not. R. Astron. Soc.}, {\bf 421}, 2277.  

Leitner S. N., Kravtsov A. V.: 2011, \journal{Astrophys. J.}, {\bf 734}, 48.  

Lemonias J. J., et al.: 2011, \journal{Astrophys. J.}, {\bf 733}, 74.  

Loh Y.-S., et al.: 2010, \journal{Mon. Not. R. Astron. Soc.}, {\bf 407}, 55.   

Lotz J. M., et al.: 2008, \journal{Astrophys. J.}, {\bf 672}, 177.  

Martig M., Bournaud F., Teyssier R., Dekel A.: 2009, \journal{Astrophys. J.}, {\bf 707}, 250.  

Martin D. C., et al.: 2005, \journal{Astrophys. J.}, {\bf 619}, L1.  

Martin D. C., et al.: 2007, \journal{Astrophys. J. Suppl. Series}, {\bf 173}, 342.  

Masters K. L., et al.: 2010, \journal{Mon. Not. R. Astron. Soc.}, {\bf 405}, 783.  

McIntosh D. H., et al.: 2014, \journal{Mon. Not. R. Astron. Soc.}, {\bf 442}, 533. 

Mendez A. J., Coil A. L., Lotz J., Salim S., Moustakas J., Simard L.: 2011, \journal{Astrophys. J.}, {\bf 736}, 110.  

Mendez A. J., et al.: 2013, \journal{Astrophys. J.}, {\bf 770}, 40.  

Meurer G. R., Heckman T. M., Calzetti D.: 1999, \journal{Astrophys. J.}, {\bf 521}, 64.  

\journal{Mon. Not. R. Astron. Soc.}, {\bf 370}, 1651.  

Moustakas J., et al.: 2013, \journal{Astrophys. J.}, {\bf 767}, 50.  

Mutch S. J., Croton D. J., Poole G. B.: 2011, \journal{Astrophys. J.}, {\bf 736}, 84.  

Nandra K., et al.: 2007, \journal{Astrophys. J.}, {\bf 660}, L11.  

Noeske K. G., et al.: 2007, \journal{Astrophys. J.}, {\bf 660}, L43.  

Peirani S., Crockett R. M., Geen S., Khochfar S., Kaviraj S., Silk J.: 2010, \journal{Mon. Not. R. Astron. Soc.}, {\bf 405}, 2327.  

Rosario D. J., et al.: 2013, \journal{Astrophys. J.}, {\bf 771}, 63.  

Saintonge A., et al.: 2011, \journal{Mon. Not. R. Astron. Soc.}, {\bf 415}, 32.  

Salim S., et al.: 2005, \journal{Astrophys. J.}, {\bf 619}, L39.  

Salim S., et al.: 2007, \journal{Astrophys. J. Suppl. Series}, {\bf 173}, 267.  

Salim S., et al.: 2009, \journal{Astrophys. J.}, {\bf 700}, 161.  

Salim S., Fang J. J., Rich R. M., Faber S. M., Thilker D. A.: 2012, \journal{Astrophys. J.}, {\bf 755}, 105.  

Salim S., Rich R. M.: 2010, \journal{Astrophys. J.}, {\bf 714}, L290.  

Schawinski K., et al.: 2007a, \journal{Astrophys. J. Suppl. Series},  {\bf 173},  512.  

Schawinski K., et al.: 2007b, \journal{Mon. Not. R. Astron. Soc.}, {\bf 381}, L74.  

Schawinski K., Virani S., Simmons B., Urry C. M., Treister E., Kaviraj S., Kushkuley B.: 2009, \journal{Astrophys. J.}, {\bf 692}, L19.  

Schawinski K., et al: 2010, \journal{Astrophys. J.},  {\bf 711},  284.  

Schiminovich D., et al..: 2010, \journal{Mon. Not. R. Astron. Soc.}, {\bf 408}, 919.  

Schiminovich D., et al.: 2007, \journal{Astrophys. J. Suppl. Series}, {\bf 173}, 315.  

Schwarz M. P.: 1984, \journal{Mon. Not. R. Astron. Soc.}, {\bf 209}, 93.  

Seibert M., et al.: 2005, \journal{Astrophys. J.}, {\bf 619}, L55.  

Silverman J. D., et al.: 2008, \journal{Astrophys. J.}, {\bf 675}, 1025.  

Smol\v{c}i\'{c} V.: 2009, \journal{Astrophys. J.}, {\bf 699}, L43. 

Speagle J. S., Steinhardt C. L., Capak P. L., Silverman J. D.: 2014, \journal{Astrophys. J. Suppl. Series}, {\bf 214}, 15.  

Stasi\'{n}ska G., et al.: 2008, \journal{Mon. Not. R. Astron. Soc.}, {\bf 391}, L29.  

Strateva I., et al.: 2001, \journal{Astron. J.}, {\bf 122}, 1861.  

Strauss M. A., et al.: 2002, \journal{Astron. J.}, {\bf 124}, 1810.  

Taylor E. N., et al.: 2011, \journal{Mon. Not. R. Astron. Soc.}, {\bf 418}, 1587.  

Thilker D. A., et al.: 2007, \journal{Astrophys. J. Suppl. Series}, {\bf 173}, 538.  

Thilker D. A., et al.: 2010, \journal{Astrophys. J.}, {\bf 714}, L171.  

Trager S. C., Faber S. M., Worthey G., Gonz\'{a}lez J. J.: 2000, \journal{Astron. J.}, {\bf 120}, 165.  

Treister E., et al.: 2009, \journal{Astrophys. J.}, {\bf 693}, 1713.  

Trump J. R., Hsu A. D., Fang J. J., Faber S. M., Koo D. C., Kocevski D. D.: 2013, \journal{Astrophys. J.}, {\bf 763}, 133.  

Vader J. P., Vigroux L.: 1991, \journal{Astron. Astrophys}, {\bf 246}, 32. 

van de Voort F., Schaye J., Booth C. M., Haas M. R., Dalla Vecchia C.: 2011, \journal{Mon. Not. R. Astron. Soc.}, {\bf 414}, 2458.  

Vasudevan R. V., Mushotzky R. F., Winter L. M., Fabian A. C.: 2009, \journal{Mon. Not. R. Astron. Soc.}, {\bf 399}, 1553.  

Vergani D., et al.: 2010, \journal{Astron. Astrophys}, {\bf 509}, AA42.  

Walker L. M., et al.: 2013, \journal{Astrophys. J.}, {\bf 775}, 129.  

Williams R. J., Quadri R. F., Franx M., van Dokkum P., Labb\'{e} I.: 2009, \journal{Astrophys. J.}, {\bf 691}, 1879.  

Wilson C. D., Scoville N.: 1991, \journal{Astrophys. J.}, {\bf 370}, 184.  

Woo J., et al.: 2013, \journal{Mon. Not. R. Astron. Soc.}, {\bf 428}, 3306.  

Wyder T. K., et al.: 2007, \journal{Astrophys. J. Suppl. Series}, {\bf 173}, 293.  

Yan R., et al.: 2011, \journal{Astrophys. J.}, {\bf 728}, 38.  

Yang Y., Zabludoff A. I., Zaritsky D., Mihos J. C.: 2008, \journal{Astrophys. J.}, {\bf 688}, 945.  

Yesuf H. M., Faber S. M., Trump J. R., Koo D. C., Fang J. J., Liu F. S., Wild V., Hayward C. C.: 2014, \journal{Astrophys. J.}, {\bf 792}, 84. 

Yi S. K., et al.: 2005, \journal{Astrophys. J.}, {\bf 619}, L111.  

York D. G., et al.: 2000, \journal{Astron. J.}, {\bf 120}, 1579.  

Young L. M., et al.: 2011, \journal{Mon. Not. R. Astron. Soc.}, {\bf 414}, 940. Agius N. K., et al.: 2013, \journal{Mon. Not. R. Astron. Soc.}, {\bf 431}, 1929.  
\endreferences

}
\end{multicols}

\end{document}